\pdfoutput=1
\documentclass[11pt]{article}

\usepackage{ACL2023}

\usepackage{times}
\usepackage{latexsym}
\usepackage{times}
\usepackage{latexsym}
\usepackage{amsmath}
\usepackage{amsfonts}
\usepackage{bbm}
\usepackage{bm}
\usepackage{graphicx}

\usepackage[T1]{fontenc}
\usepackage[utf8]{inputenc}
\usepackage{multirow}
\usepackage{microtype}

\usepackage{inconsolata}
\usepackage{microtype}
\usepackage{ulem}
\usepackage{tabularx}

\newcommand{\ourmodel}{\textsc{KLEVER}}

\title{Improving Items and Contexts Understanding with Descriptive Graph for Conversational Recommendation}

\author{Huy Dao \\ FPT Software - AIC \\ Hanoi, Vietnam \\ huydq44@fsoft.com.vn
        \And  Dung D. Le \\ VinUniversity \\Hanoi, Vietnam \\ dung.ld@vinuni.edu.vn \And
        Cuong Chu \\ Max Planck Institute for Informatics \\ Germany \\ cxchu@mpi-inf.mpg.de }

\begin{document}
{\makeatletter\acl@finalcopytrue
  \maketitle
}

\begin{abstract}
State-of-the-art methods on conversational recommender systems (CRS) leverage external knowledge to enhance both items' and contextual words' representations to achieve high quality recommendations and responses generation. However, the representations of the items and words are usually modeled in two separated semantic spaces, which leads to misalignment issue between them. Consequently, this will cause the CRS to only achieve a sub-optimal ranking performance, especially when there is a lack of sufficient information from the user's input. To address limitations of previous works, we propose a new CRS framework \ourmodel, which jointly models items and their associated contextual words in the same semantic space. Particularly, we construct an \textit{item descriptive graph} from the rich items' textual features, such as item description and categories. Based on the constructed descriptive graph, \ourmodel{} jointly learns the embeddings of the words and items, towards enhancing both recommender and dialog generation modules. Extensive experiments on benchmarking CRS dataset demonstrate that \ourmodel{} achieves superior performance, especially when the information from the users' responses is lacking.

\end{abstract}

\section{Introduction}

\begin{figure}
    \centering
    \includegraphics[scale=0.75]{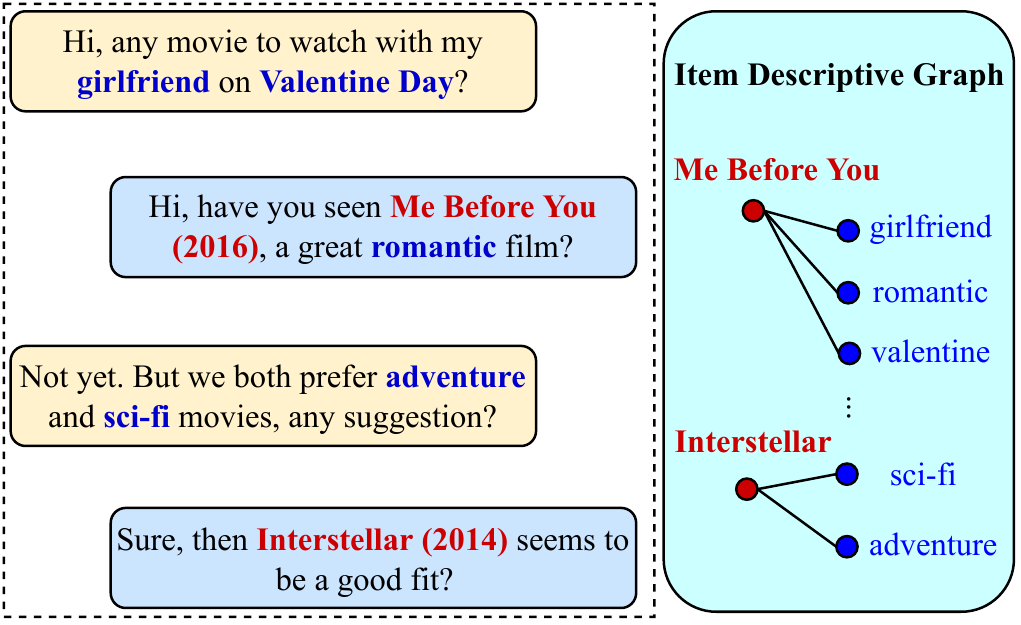}
    \caption{The three words \textit{girlfriend}, \textit{valentine}, \textit{romantic} and the item \textbf{Me Before You} express some common knowledge. Similar to the two words \textit{adventure}, \textit{sci-fi} and the item \textbf{Interstellar}. An \textit{item descriptive graph} connects these information and reduces false alignments between items and words.}
    % \vspace{-0.1in}
    \label{fig:problem}
\end{figure}

{\bf Motivation and Problem.} Recently, recommender systems \cite{gcmc,rs_survey, star_gcn,ncf,ngcf,light_gcn,le2017indexable,le2021efficient} have been investigated extensively due to their practical benefits for industrial applications. Such information retrieval systems provide personalized recommendations to the users based on their historical interactions such as recorded ratings or clicking history. However, conventional recommender systems suffer from the cold-start problem wherein the systems need to recommend items to new users or the user's interactions are very scarce \cite{survey}. Besides, it is hard for static recommender systems to capture online user's preferences, since the user's interests are often dynamic and vary over time \cite{survey}. For those reasons, conversational recommender systems (CRSs) \cite{tcrm, crm, kgsf, survey} have gained considerable attention from both academic researchers and industry practitioners, thanks to their ability to offer interactive experience and recommend suitable items on the fly to the users. The goals of such systems are to produce relevant recommendations by interactively asking clarifying questions \cite{qrec, ear, scpr,kbqg} and to generate human-like and informative responses \cite{redial, kbrd, kgsf, crwalker}.

A desirable quality of CRS frameworks is that the systems could produce appropriate recommendations by only comprehending some indicate words from the conversations with the users. For example, in the figure \ref{fig:problem}, the words ``girlfriend, valentine, romantic'' possibly correspond to the item \texttt{Me Before You (2016)}, while ``adventure, sci-fi'' are more likely suitable for \texttt{Interstellar (2014)}. Understanding such relationships between items and words is especially useful in many cases when the users may not be familiar with available items and only express their preferences by mentioning some descriptive words. However, state-of-the-art approaches \cite{kgsf, revcore} suffer from the problem of misalignment between those two kinds of information since word and item representations in their proposed frameworks are inherently modeled in two separated semantic spaces.

\noindent {\bf Approach.} 
We argue that the misalignment between item and context representations can be addressed by jointly learning them in a common semantic space. Previous CRS frameworks suffer from the misalignment problem due to the following two reasons: (1) do not utilize the contextual descriptive features to enrich the understanding of the items and (2) lacking of an effective mechanism to jointly learn word and item embeddings in the same semantic space. To this end, we propose a novel CRS framework called \ourmodel{} ({\it \underline{K}now\underline{L}edge \underline{E}nhanced con\underline{VE}rsational \underline{R}ecommender System}) which jointly learn item and word representations in the same semantic space. To facilitate our proposed framework, we introduce (1) an item descriptive graph, constructed by extracting descriptive terms from entities' textual features (such as item categories, item reviews, entity descriptions). This heterogeneous graph captures semantic relationships between items and their descriptive words and (2) a self-supervised learning setting to jointly learn word and item representations. The sets of learned representations are then used to enhance the performance of both recommendation and dialog generation modules. 

\noindent {\bf Contributions and Organization.} Our contributions can be summarized as follows: Firstly, we introduce a novel item descriptive graph that serves as a prior resource for enhancing the accuracy of items' and words' representations alignment. To the best of our knowledge, this is the first work that an item descriptive graph is introduced to better learn the representations for items and improve the performance of conversational recommendations. Secondly, based on the constructed graph, we learn the embeddings for items and contextual words to optimize for a self-supervised link prediction task. These embeddings are used for two downstream tasks: item recommendation and response generation. Besides, we also introduce a bag-of-words loss based on the connectivity of our item descriptive graph to promote the model to generate relevant words with entities mentioned in the conversation. {\it Thirdly,} we conduct extensive experiments on a public CRS dataset to demonstrate the superiority of \ourmodel{} compared with state-of-the-art competitors in Sections \ref{sec:experimental_setup} and \ref{sec:experimental_result}. Our detailed analysis further shows its advantages in the cold-start setting where the CRS needs to predict suitable recommended items using only the word-based information. For completeness, we discuss related work in Section \ref{sec:related_work} and conclude in Section \ref{sec:conclusion}. 

\section{Related Work}
\label{sec:related_work}

\begin{figure*}
    \centering
    \includegraphics[scale=0.35]{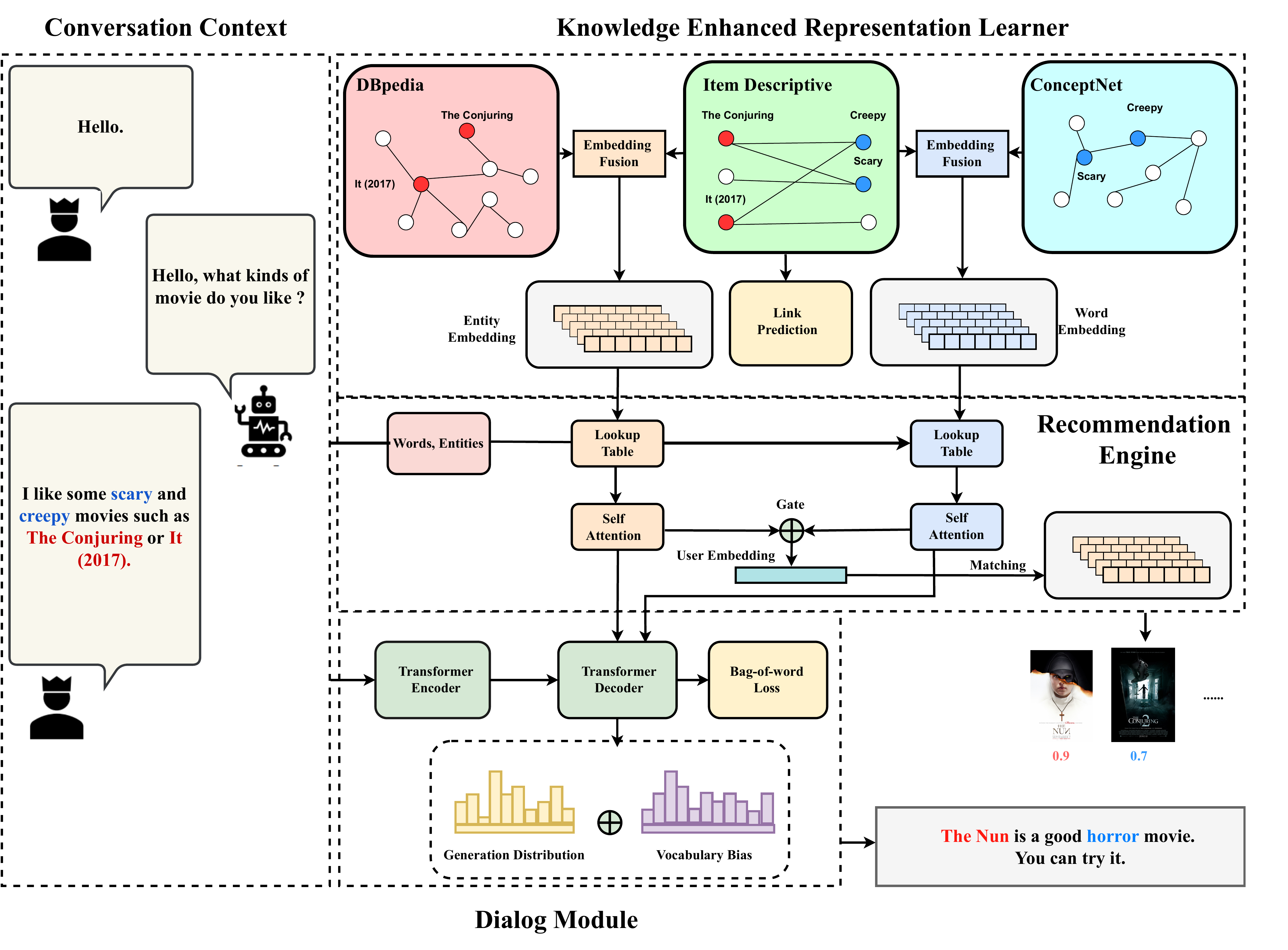}
    \caption{The overall framework of our model.}
    \label{fig:model}
    % \vspace{-0.1in}
\end{figure*}

\paragraph{\bf{Attribute-based CRS}} 

In this setting, CRS models focus on asking clarifying questions on the item attributes. There are several works following this line of research including reinforcement learning based techniques \cite{crm, ear, unicorn, conts}, generalized binary search \cite{qrec}, graph-based approaches \cite{memory_graph, scpr, kbqg, adapt}, memory network \cite{saur}, adversarial learning \cite{crsal} and multi-armed bandit algorithms \cite{tcrm, conts}.  
Another line of research focuses on balancing exploration (\textit{i.e} asking questions) and exploitation (\textit{i.e} recommending items) trade-off for cold-start users \cite{survey, tcrm, conts}. Though attribute-based CRS models are more controllable in industrial applications; however, such CRS systems lack of the ability to naturally interact with the users, which may lead to undesirable user experience.  

\paragraph{\bf{Dialog-based CRS}} Recently, dialog-based CRS models \cite{redial, kbrd, deepcrs, botplay, multi_type, kgsf, explainable, inspired, tg_redial, ntrd, revcore} have been extensively investigated due to their flexibility and interactiveness. \citet{redial} propose a benchmark dataset called REDIAL which is collected from Amazon Mechanical Turk (AMT). The dataset consists of a large amount human conversations in movie recommendation scenarios. State-of-the-art dialog-based CRS models \cite{kbrd,kecrs, kgsf,revcore} propose to incorporate domain knowledge to enhance the semantic meaning of the conversation.  
\citet{kgsf} leverage two knowledge graphs, \textit{i.e} Dbpedia \cite{dbpedia} and ConcepNet \cite{conceptnet} to connect related entities and words respectively and learn those two pieces of information in two separated vector spaces. Moreover, they leverage a Mutual Information Maximization (MIM) objective \cite{dgi, mutual} to align word and entity representations that express the same knowledge.

Our work lies in the research of dialog-based CRS. In contrast to previous works \cite{kgsf} \cite{revcore}, we propose to jointly learn item and context representations in the same semantic space by using the proposed item descriptive graph and a self-supervised objective function. 

\section{Preliminaries}

For convenience, we denote $\mathcal{I}, \mathcal{V}$ as set of items and the vocabulary respectively. We also denote the context dialog as $\textbf{s} = \{ s_1, s_2, ...., s_t \}$ where $s_i$ is the sentence at $i$-th turn and $s_{i} = \{t_{i,1}, t_{i,2}, ..., t_{i,N} \}$ where $t_{i,j}$ is the $j$-th token at the $i$-th sentence. Given the context $\mathbf{s}$, CRS models try to produce proper items and generate natural responses based on extracted information from the context. Formally, at the $t$-th conversation turn, the recommendation engine retrieves a set of candidate items $\mathcal{I}_{t+1}$ from the entire set $\mathcal{I}$ while the dialog module generates the next utterance $s_{t+1}$ to respond the user.

Following \cite{kbrd,kgsf}, we adopt Dbpedia \cite{dbpedia} and ConceptNet \cite{conceptnet} \cite{arabshahi2021} to build the item-oriented and word-oriented knowledge graphs respectively. Then we utilize Relational Graph Convolutional Network (RGCN) \cite{rgcn} and Graph Convolutional Network (GCN) \cite{gcn} to learn entity and word representations respectively. Finally, we obtain an embedding $\textbf{e}_{u} \in \mathbb{R}^{d} $ for each entity $u$ and $\textbf{e}_{w} \in \mathbb{R}^{d}$ for each word $w$ where $d$ is the dimensionality of those representations.

\section{Methodology}
\label{sec:methodology}

In this section, we describe our proposed framework to the CRS task. We firstly describe how we construct the proposed item descriptive graph which consists of nodes of items and descriptive words. Each item node has connections to its corresponding words and vice versa. Based on the constructed graph, we propose a self-supervised objective function to jointly learn item and word representations in the same semantic space. Finally, we introduce the recommendation engine and the dialog module based on the learned representations. Figure \ref{fig:model} depicts our proposed CRS framework.

\subsection{Item Descriptive Graph}

For handling the misalignment between words and items, we propose an Item Descriptive Graph (IDG for short) where in each item is represented by a set of descriptive words. These words are directly retrieved from item's meta information such as item categories and associated tags, as well as other textual features such as user reviews and entity descriptions by using a simple linguistics approach (i.e removing stopwords and only considering words whose frequencies appear more than a certain threshold value $m$). In the end, only important tags (categories) and top-$k$ frequent words are used to form the representative set of each item. Our proposed IDG have several advantages: (1) it provides the prior knowledge for handling the misalignment between items and their contextual words based on rich textual features, and (2) similar items may share a similar set of descriptive words. Hence leveraging the constructed graph might help the model capture meaningful properties of the items. We depict the item descriptive graph in figure \ref{fig:problem}.

\subsection{Joint Learning Words and Items Representations}

In contrast to previous works \cite{kgsf} \cite{revcore}, our method directly handle the aforementioned misalignment by modeling both items and words representations in the same semantic space. Specifically, with the constructed item descriptive graph, we utilize a GCN model \cite{gcn} to jointly learn the item and word embeddings as follows:

\begin{equation}
    \begin{aligned}
        \hat{\textbf{e}}_{u}^{(l)} &= \sigma(\sum_{w \in N(u)} \hat{\textbf{W}}^{(l)} \hat{\textbf{e}}_{w}^{(l-1)} + \hat{\textbf{W}}_{0} \hat{\textbf{e}}_{u}^{(l-1)}) \\ 
        \hat{\textbf{e}}_{w}^{(l)} &= \sigma(\sum_{u \in N(w)} \hat{\textbf{W}}^{(l)} \hat{\textbf{e}}_{u}^{(l-1)} + \hat{\textbf{W}}_{0} \hat{\textbf{e}}_{w}^{(l-1)}) \\ 
    \end{aligned}
\end{equation}
where $\hat{\textbf{e}}_{u}^{(l)}, \hat{\textbf{e}}_{w}^{(l)} \in \mathbb{R}^{d}$ are enhanced representations and $N(u), N(w)$ are set of neighbors for item $u$ and word $w$ respectively. $\hat{\textbf{W}}^{(l)}, \hat{\textbf{W}}^{l}_{0}$ are shared weight matrices for both entities and words at the $l$-th layer. Besides, we adopt Leaky ReLU \cite{leaky_relu} as the non-linear activation function.

\medskip
\paragraph{Link-prediction Loss:} To effectively guide the representation learning process on the item descriptive graph, we propose a self-supervised learning approach based on the link prediction task.  Specifically, we learn the embeddings $\hat{\textbf{e}}^{(l)}_{w}$ and $\hat{\textbf{e}}^{(l)}_{u}$ to predict whether existing a link between the word $w$ and the entity $u$ with the probability computed as follows:
\begin{equation}
    \begin{aligned}
     \text{p}_{w,u} = \sigma \left((\hat{\textbf{e}}^{(l)}_{w})^{T}\hat{\textbf{e}}^{(l)}_{u} \right)
    \end{aligned}
\end{equation}
where $\sigma$ denotes for the sigmoid function. Following \cite{word2vec, rgcn}, we utilize negative sampling to train the link prediction loss:

\begin{equation*}
    \begin{aligned}
        L_{link} =  - \frac{1}{N} \sum_{(w,u) \in \mathcal{E}^{+} \bigcup \mathcal{E}^{-}} \mathbbm{1}[(w,u) \in \mathcal{E}^{+}] \log(\text{p}_{w,u})  \\ 
        + \mathbbm{1}[(w,u) \in \mathcal{E}^{-}] (1 - \log(\text{p}_{w,u})) %\hspace{0.7cm} (3)
    \end{aligned}
\end{equation*}
where $\mathcal{E}^{+}, \mathcal{E}^{-}$ are sets of positive and negative examples respectively. Hence, $N = |\mathcal{E}^{+}| + |\mathcal{E}^{-}|$ is the total number of training examples. 

\medskip
\paragraph{Embedding Fusion:} 

Finally, we obtain the final representations for words and items by fusing embedding vectors learned by the item descriptive, item-oriented and word-oriented knowledge graphs respectively. Given a word $w$ and an item $u$, we obtain the final embeddings for item $u$ and word $w$ using the following formulations.

\begin{equation}
    \begin{aligned}
        \textbf{h}_{u} &= \textbf{W}_{u} [\textbf{e}_{u}, \hat{\textbf{e}}_{u}] + \textbf{b}_{u} \\
        \textbf{h}_{w} &= \textbf{W}_{w} [\textbf{e}_{w}, \hat{\textbf{e}}_{w}] + \textbf{b}_{w} \\
    \end{aligned}
\end{equation}

\noindent where $\textbf{h}_{u}, \textbf{h}_{w} \in \mathbb{R}^{d}$ are fused representations for entity $u$ and word $w$. $\textbf{W}_{u}, \textbf{W}_{w} \in \mathbb{R}^{2d \times d}$ and $\textbf{b}_{u}, \textbf{b}_{w} \in \mathbb{R}^{d}$ are learnable weight matrices and biases for items and words respectively. 

\paragraph{Misalignment Handling:} Intuitively, representations of items and their contextual words (e.g \texttt{Ironman} and \textit{super hero} or \texttt{The Conjuring} and \textit{horror} ) should be close to each other in the embedding space. In the item descriptive graph, each item has several connections to its contextual words that are extracted from rich item-side information such as item categories, tags, keywords. By adapting a GNN model \cite{gcn} which acts as a smoothing operator \cite{smoothing}, the joint learner is able to produce similar representations for each node and its neighbor nodes in the graph. Besides, items that share a common set of contextual words may also have similar representations, which improves the quality of the learned embedding vectors.

\subsection{Knowledge-enhanced Recommendation Engine}
Given a conversation context $\mathbf{s} = \{ s_1, s_2, ...., s_t \}$, we first extract all words and entities, then lookup for their fused embeddings learnt from the previous step.  
Via a gating network, the user preference $\textbf{u}$ is then defined as a combination of word and entity representations as follows:

\begin{equation}
    \begin{aligned}
     \textbf{u} &= \beta * \textbf{p}_u + (1 -\beta) *\textbf{p}_w \\
     \beta &= \sigma(\textbf{W}_{\text{gate}}[\textbf{p}_u, \textbf{p}_w] + \textbf{b}_{\text{gate}})
    \end{aligned}
\end{equation}
where $\textbf{p}_u$ and $\textbf{p}_w$ are vector representations for entity and context information respectively. We compute those embedding vectors by using the self-attention mechanism \cite{kgsf}.

Finally, given the user preference vector $\textbf{u}$, the probability that item $i$ is recommended is the dot product between user preference vector $\textbf{u}$ and the fused item embedding $\textbf{h}_{i}$.
\begin{equation}
    \begin{aligned}
    \text{P}_{rec}(i) = \text{Softmax}(\textbf{u}^{T}\textbf{h}_{i})
    \end{aligned}
\end{equation}
To train the recommendation engine, we optimize the cross entropy loss between our model prediction and the target item.

\begin{equation*}
    \begin{aligned}
    L_{rec} = - {\sum_{s \in \mathcal{S}}} \sum_{i \in \mathcal{I}} &(1 - y_{i}^{s}) * \log(1 - \text{P}^{s}_{rec}(i)) \\
    &+ y^{s}_{i} \log \text{P}^{s}_{rec}(i) 
    + \lambda_{1} * L_{link} %\hspace{1.25cm} (6)
    \end{aligned}
\end{equation*}
where $\mathcal{S}$ is set of all conversations and $y^s_{i}$ is the label of item $i$ at the conversation $s$. We optimize the recommendation loss and the link prediction loss jointly whereas $\lambda$ is a weighted parameter.

\subsection{Knowledge-enhanced Dialog Module}

For response generation module, we adopt Transformer \cite{transformer} as the main architecture. Given a conversation context, we utilize Transformer Encoder to encode the context sequence. At the $j$-th decoding step, we feed the context features $\textbf{x}_{all}$ as well as embeddings of groundtruth tokens before the $j$-th position $\{y_{1}, y_{2},..., y_{j-1}\}$ into the Transformer Decoder to obtain a hidden vector $\textbf{s}_{j} \in \mathbb{R}^{d_{gen}}$ that represents information for predicting the $j$-th token $y_j$. 

\paragraph{\bf{Bag-of-words Loss:}} 

To promote the model to generate relevant words with items mentioned in the conversation, we design a novel sentence-level bag-of-words loss based on the connectivity of our constructed item descriptive graph. Firstly, we compute a vector $\textbf{a}_{j} \in \mathbb{R}^{|\mathcal{V}|}$ representing the predicted scores at the $j$-th position in the response as follows:
\begin{equation}
    \begin{aligned}
        \textbf{a}_{j} = \textbf{W}_{bow}[\textbf{s}_{j}, \textbf{p}_{u}, \textbf{p}_{w}] + \textbf{b}_{bow}
    \end{aligned}
\end{equation}
where $\textbf{W}_{bow} \in \mathbb{R}^{d_{gen} + 2 * d \times |\mathcal{V}|}, \textbf{b}_{bow} \in \mathbb{R}^{|\mathcal{V}|}$ are learnable parameters. We define a probability distribution $\text{P}_{bow}$ whose each element represents how likely each word $w$ in the vocabulary $\mathcal{V}$ appears in the generated sentence regardless 
of the position in the sentence as follows.
\begin{equation}
    \begin{aligned}
        \text{P}_{bow} = \sigma\left(\sum_{j=1}^{N}\textbf{a}_{j}\right)
    \end{aligned}
\end{equation}
where $N$ is the number of words in the response and $\sigma$ is the sigmoid function.

Then we optimize the bag-of-word objective by using the following loss function. 

\begin{equation}
    \begin{aligned}
        L_{bow} = - \sum_{u \in \textbf{s}} \sum_{w \in \mathcal{N}_{1-hop}(u)} \log(\text{P}_{bow}(w))
    \end{aligned}
\end{equation}
where $\mathcal{N}_{1-hop}(u)$ is the set of 1-hop words of entity $u$ in the conversation context $s$. Finally, we compute the probability distribution at the $j$-th token by the following formulation.
\begin{equation}
    \begin{aligned}
     \text{Pr}(y_j) = \text{Pr}_{1}(y_j|\textbf{s}_{j}) +  \text{Pr}_{2}(y_j|\textbf{s}_{j}, \mathcal{G}_{1}, \mathcal{G}_{2}, \mathcal{G}_{3} ) \\
     + \text{Pr}_{3}(y_j|\textbf{s}_{j}, \mathcal{G}_{1}, \mathcal{G}_{2}, \mathcal{G}_{3} )
    \end{aligned}
\end{equation}
where $\text{Pr}_{1}(.)$ is the generative probability computed by the output of the Transformer Decoder, $\text{Pr}(.)_{2}$ is the copy probability implemented by the standard copy mechanism \cite{copy} and $\text{Pr}_{3}(.)$ is the knowledge-guided bag-of-words probability and is computed by applying the Softmax function over the predicted vector $\textbf{a}_{j}$.

To train the dialog module, we optimize the cross-entropy loss of ground truth responses and the bag-of-words loss jointly. The final generation loss function is defined as follows.
\begin{equation*}
    % \begin{aligned}
    L_{gen} = - \frac{1}{N} \sum_{j=1}^{N} \log(\text{Pr}(y_{j}|\textbf{s}, y_{1}, ...y_{j-1})) +  \lambda_{2} L_{bow}
    % \end{aligned}
\end{equation*}
where $\textbf{y} = \{y_{1}, y_{2}, ..., y_{N}\}$ is the corresponding ground truth response of the conversation context $\textbf{s}$ and $\lambda_{2}$ is the weighted hyperparameter for the bag-of-words loss. 

\section{Experimental Setup}
\label{sec:experimental_setup}
\subsection{Dataset}

We conduct all experiments on the REDIAL dataset, a recent benchmark for CRS models introduced in \cite{redial}. For entity textual features to construct the item descriptive graph, following \citet{revcore}, we crawl movie genres, movie keyplots, user reviews and entity descriptions from IMDB website \footnote{\url{https://www.imdb.com/}}.
The detailed statistics of Redial dataset, our constructed item descriptive graph and the model’s implementations can be seen in table 1, 2 at Appendix A.1

%%%%%%%%%%%%%%%%%%%%%%%%%%%%%%%%%%%%%
%%%%%%%%%%%%%%%%%%%%%%%%%%%%%%%%%%%%%
\begin{table}[t]
\centering
\caption{ $\mathbf{R}$@$k$ for the recommendation task.}
\scalebox{0.7}{
\begin{tabular}{lc|c|c|c|c|c|c|}
  \hline
  \multirow{2}{*}{Model} 
      & \multicolumn{3}{c}{All Data} 
          & \multicolumn{3}{|c|}{Cold Start} \\             \cline{2-7}
  & \textbf{R@1} & \textbf{R@10} & \textbf{R@50} & \textbf{R@1} & \textbf{R@10} & \textbf{R@50} \\  \hline
Redial & 0.024 & 0.140 & 0.320 & 0.021 & 0.075 & 0.201 \\     
KBRD & 0.031 & 0.150 & 0.336 & 0.026 & 0.085 & 0.242 \\   
KECRS & 0.025 & 0.153 & 0.349 & 0.032 & 0.148 & 0.327\\ 
KGSF & 0.036 & 0.182 & 0.373 & 0.036 & 0.168 & 0.368 \\ 
RevCore & 0.037 & 0.187 & 0.380 & 0.033 & 0.168 & 0.365 \\  
KLEVER & \textbf{0.041} & \textbf{0.203} & \textbf{0.383}& \textbf{0.049} & \textbf{0.184} & \textbf{0.369} \\      \hline
KLEVER - L & 0.037 & 0.199 & 0.374 & 0.047 & 0.177 & 0.361\\
KLEVER - IDG & 0.033 & 0.173 & 0.346 & 0.035 & 0.168 & 0.359\\
KLEVER - KG & 0.020 & 0.100 & 0.237 & 0.019 & 0.089 & 0.245\\\hline
% KLEVER - L & 0.037 & 0.199 & 0.374 & 0.047 & 0.177 & 0.361\\
% KLEVER - F & 0.033 & 0.173 & 0.346 & 0.035 & 0.168 & 0.359\\
% KLEVER - KG & 0.020 & 0.100 & 0.237 & 0.019 & 0.089 & 0.245\\\hline
\end{tabular}}
% \vspace{-0.1in}
% \vspace{-0.2in}
\label{tab:rec_results}
\end{table}
%%%%%%%%%%%%%%%%%%%%%%%%%%%%%%%%%%%%%
%%%%%%%%%%%%%%%%%%%%%%%%%%%%%%%%%%%%%

\subsection{Baseline Methods}

We compare our CRS framework denoted as \textbf{KLEVER} with several representative baseline approaches: \textbf{REDIAL} \cite{redial}, which utilizes a sentiment-aware auto-encoder \cite{auto} as the recommendation model. \textbf{KBRD} \cite{kbrd}, a model utilizes a knowledge graph based on DBpedia \citet{dbpedia} to enhance entity representations. \textbf{KECRS} \cite{kecrs}, which adopts a domain-specific knowledge graph based on The Movie Database (TMDB) \footnote{https://www.themoviedb.org/}.  \textbf{KGSF} \cite{kgsf}, which leverages two KGs, i.e ConceptNet and DBpedia, and introduces the Mutual Information Maximization objective to bridge the gap between concept and item representations. \textbf{RevCore} \cite{revcore}, which incorporates user's reviews to enhance the semantic of the conversation. Besides, we also denote \textbf{KLEVER - L}, \textbf{KLEVER - Bow}, \textbf{KLEVER - IDG}, \textbf{KLEVER - KG} as variants of our model without link-prediction loss, bag-of-words loss, item descriptive graph and knowledge graphs (i.e. DBpedia, ConceptNet) respectively.    

\subsection{Evaluation Metrics}

For \textit{recommendation task}, we use Recall@k (originally defined in \cite{redial} and also used in CRS methods \cite{kbrd,kgsf}), denoted as \textbf{R@k} ($k = 1, 10, 50$) which checks whether the top-k recommended items contain the ground-truth item. Besides, we also simulate the \textbf{cold-start scenario} in CRS, \textit{i.e} we only consider the test examples without any mentioned items in the conversation context. 

For the \textit{generation task}, we assess the generated responses in both automated and manual manners. For automated evaluation, we utilize Distinct N-gram ($N = 2, 3, 4$) \cite{diverse, kgsf} to measure the diversity of generated sentences. For manual evaluation, we randomly select 50 conversations and responses generated by \ourmodel{} and baseline models. We invite three annotators to score the responses in two aspects, Fluency and Informativeness. The range of score is 1 to 3. The final performance is calculated using the average scores of all annotators. The inter-annotator agreement is measured by Fleiss’ Kappa \cite{kappa}. Detailed instruction for manual evaluation can be found at Appendix A.2. %\ref{sec:instruct-manual-eval}.

\section{Experimental Results}
\label{sec:experimental_result}
\subsection{Recommendation Performance}

\paragraph{\bf{Evaluation on All Data Setting:}} 
As can be seen in table \ref{tab:rec_results},  
our model outperforms all baseline methods in all metrics and achieves the state-of-the-art performance. Noticeably, our model significantly performs better than KGSF (+13.88\% R@1, +11.22\% R@10, +2.68\% R@50). The reason is possibly that KGSF and RevCore inherently capture words and items in two separated semantic spaces, which lead to the mismatch between the two kinds of signal. While our proposed model is able to jointly model them in a common semantic space and alleviate the mismatch by aligning items with their descriptive words extracted from items' textual features.
\paragraph{\bf{Evaluation on Cold Start Setting:}} 

As one can see in table \ref{tab:rec_results}, 
our model also outperforms all baseline methods in all metrics, superior over KGSF (+36.11\% R@1, +9.52\% R@10) and RevCore (+48.48\% R@1, +9.52\% R@10). 
Both KGSF and RevCore perform poorly on the cold-start setting since they suffer from the misalignment between words and items. On the other hand, our proposed model can provide an effective alignment between those two kinds of information. Therefore, when the user mentions some indicative words in the conversation, the model may get more evidence to recommend relevant items associated with these mentioned words.

\subsection{Generation Performance}

\begin{table}
\centering
\caption{Performance on the generation task.}
\begin{tabular}{lc|c|c|c}
\hline
\textbf{Models} & \textbf{Dist-2} & \textbf{Dist-3} & \textbf{Dist-4}\\
\hline
Redial & 0.225 & 0.236 & 0.228 \\
KBRD & 0.263 & 0.368 & 0.423  \\
KECRS & 0.286 & 0.392 & 0.451 \\ 
KGSF & 0.364 & 0.517 & 0.605  \\ 
RevCore & 0.391 & 0.568 & 0.667\\ 
KLEVER & \textbf{0.427} & \textbf{0.622} & \textbf{0.743}  \\\hline
KLEVER - Bow  & 0.391 & 0.578 & 0.701  \\
KLEVER - IDG & 0.380 & 0.524 & 0.594  \\\hline
\end{tabular}

\label{tab:gen_results}
\end{table}

\paragraph{\bf{Automatic Evaluation:}} Table \ref{tab:gen_results} shows the generation performance of CRS models. 
Compared with the baseline models, our proposed model is consistently better in all evaluation metrics. Noticeably, our model outperform the RevCore model (+9.21\% Dist2, +9.50\% Dist-3, +11.39\% Dist-4) and achieves the state-of-the-art performance. We hypothesize that our new bag-of-words objective provides additional guidance for the dialog module to generate sentences not only from ground-truth sequences but also from rich entities textual features; therefore, it may improve the diversity of the generated responses.

\begin{table}
\centering
\caption{ Human evaluation on the generation task.}
\begin{tabular}{lc|c|c|c}
\hline
\textbf{Models} & \textbf{\small Fluency} & \textbf{\small Informativeness} & \textbf{Kappa} \\
\hline
KECRS & 2.86 & 1.25 & 0.82 \\
KGSF & 2.71 & 1.55 & 0.74  \\ 
RevCore & 2.79 & 1.40 & \textbf{0.86}  \\
KLEVER & \textbf{2.90} & \textbf{1.85} & 0.76   \\\hline
Human & 2.84 & 2.25  & 0.65  \\ \hline

\end{tabular}

\label{tab:human_eval}
\vspace{-0.2cm}
\end{table}

\paragraph{\bf{Human Evaluation:}} Table \ref{tab:human_eval} summarizes performance of CRS models by human evaluation. For fluency, all considered models achieve similar and high scores. The reason is possibly from the fact that those models tend to generate short and safe responses. 
On the other hand, for informativeness, our model significantly outperforms all baseline methods. By leveraging the richness from the item descriptive graph with our proposed bag-of-words objective, rather than producing safe and short sentences, our model is able to generate relevant information, especially in such cold-start cases, which improves the informativeness of the responses.

\subsection{Ablation Study}

\paragraph{\bf{Recommendation Task:}} 
As can be seen in table \ref{tab:rec_results}, compared to our best model, removing either the self-supervised objective function (i.e. link prediction loss) or item descriptive graph leads to a sharp decreasing on the recommendation accuracy in both the all-data and cold-start settings. We hypothesize that the link-prediction loss helps the model to infer potential edges between items and their descriptive words, which may be missed in our constructed graph. The result also shows that the item descriptive graph are crucial to handle cold-start cases where the model need to effectively align the contextual words to potential items to produce appropriate recommendations to the users. 

\paragraph{\bf{Generation Task:}} 
Table \ref{tab:gen_results} demonstrates that all the components in \ourmodel{} are significantly useful on the response generation task. 
The bag-of-words loss is able to guide the model in generating relevant words that may not belong to the ground-truth sentences but are handled by rich and diverse textual features. On the other hand, the proposed joint learning process through the item descriptive graph is able to handle the false alignments between word and entity representations, not only enhances the recommendation performance but it also improves the quality of the generated responses.

\subsection{Discussion}

\paragraph{\bf{Item Descriptive Graph improves MIM.}} To demonstrate the effectiveness of our contribution, we also incorporate the item descriptive graph into the KGSF model. Specifically, we only utilize the MIM objective to align entities with their corresponding descriptive words based on the graph. Table \ref{tab:we_study} shows that by incorporating the item descriptive graph, the KGSF model is able to achieve better performance on the recommendation task. The reason is that our proposed item graph alleviates noisy alignments between words and entities, therefore, improves the quality of entity and word representations. Noticeably, our proposal \ourmodel{} still significantly performs better than the KGSF + IDG model. 
We also conduct an analysis on the number of descriptive words per each item in the item descriptive graph, which can be found at Appendix A.3. %\ref{sec:exp-num-des-words}.

\begin{table}
\centering
\caption{KGSF with Item Descriptive Graph on the recommendation task.}
\begin{tabular}{lc|c|c|c}
\hline
\textbf{Model} & \textbf{R@1} & \textbf{R@10} & \textbf{R@50}\\
\hline
KGSF  & 0.036 & 0.182 & 0.373  \\
KGSF + IDG  & 0.038 & 0.183 & 
0.381  \\ 
KLEVER & \textbf{0.041} & \textbf{0.203} & \textbf{0.383} \\ \hline
\end{tabular}
\label{tab:we_study}
%\vspace{-0.3cm}
\end{table}

\begin{figure}
    \centering
    \includegraphics[scale=0.33]{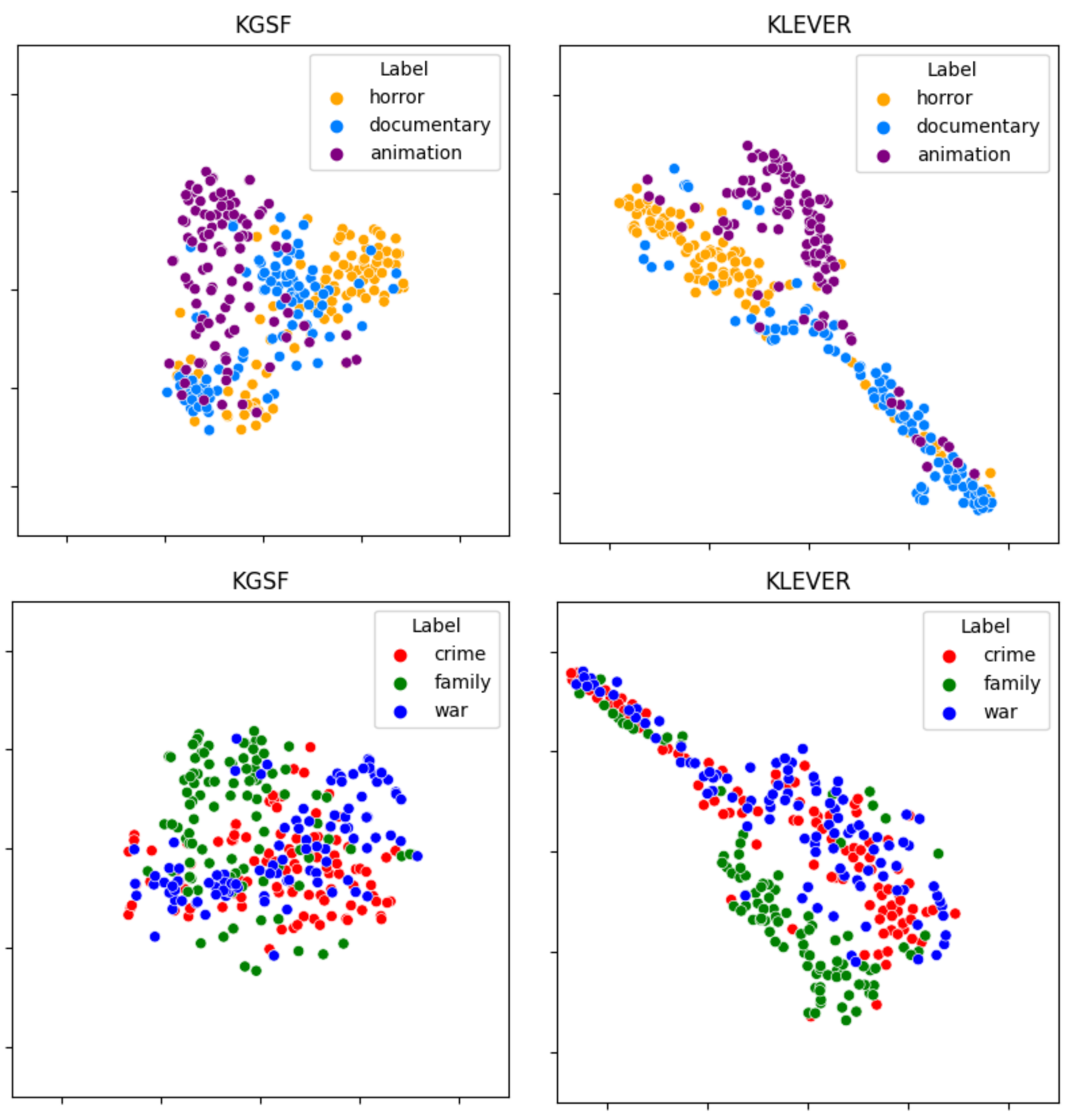}\    \caption{Item embedding vectors learned by KGSF (left) and KLEVER (right). We use TSNE \cite{tsne} to produce the low dimensional visualizations. Each row illustrates item representations from 3 distinct categories.}
        \label{fig:embed_visual}
    \vspace{-0.3cm}
\end{figure}

\paragraph{Case Study I: Embedding Visualization} We visualize item embedding vectors learned by KGSF \cite{kgsf} and our proposed model to demonstrate that join learning words and items can lead to more meaningful representations. Figure \ref{fig:embed_visual} depicts the learned item embeddings according to some randomly chosen item categories. We can see that item representations produced by KLEVER are more separable and meaningful than those learned by KGSF. This is reasonable since our model directly align items with categories by using prior knowledge from the item descriptive graph.

\paragraph{\bf{Case Study II: Interactive Dialog}}
Table \ref{tab:case_study} shows an anecdotal example of a cold-start conversation when no item is mentioned in the beginning of conversation. As one can see, KLEVER is able to generate more informative and meaningful responses than other baseline methods. Other examples can be also found at Appendix A.4. 

\begin{table}
\centering
\caption{ Case Study. A cold-start example on movie recommendation. Human responses are ground-truth.}
\scalebox{0.82}{
\begin{tabularx}{1.2\columnwidth}{lX}
\hline
& \textbf{Conversation Example} \\
\hline
\textbf{User:}  & Hello. Do you know any good \textcolor{blue}{horror} movies? \\\hline
\textbf{Human:} & \textcolor{red}{Rings (2017)} \\
% \textbf{KECRS:} & Hello ! What kind of movies do you like ? \\
\textbf{KGSF:}  &  Hello, how are you? \\ 
\textbf{Revcore:} & Hello! What kind of movies do you like \\
\textbf{KLEVER:}  & I would recommend \textcolor{red}{The Conjuring (2013)} \\\hline
\textbf{User:}  & What is \textcolor{red}{Rings (2017)} about?\\\hline
\textbf{Human:} & It is \textcolor{blue}{horror} movie where the little girl fears all as a \textcolor{blue}{ghost} coming out of the well\\
% \textbf{KECRS:} & I haven't seen that one yet . \\
\textbf{KGSF:}  & I would recommend you to watch it  \\ 
\textbf{Revcore:} &  I also recommend \textcolor{red}{Hellraiser (1987)} \\
\textbf{KLEVER:}  & I'm not sure if you like \textcolor{blue}{ghost} movies .\\\hline
\textbf{User:}  & I think I've seen part of it\\\hline
\textbf{Human:} & Oh nice why don't you see the full movie\\
% \textbf{KECRS:} & I haven't seen that one . \\
\textbf{KGSF:}  & I would recommend you to watch it. It's a bit of Stephen King. \\ 
\textbf{Revcore:} &  I also recommend \textcolor{red}{The Babadook (2014)} \\
\textbf{KLEVER:}  &  I would recommend you \textcolor{red}{68 Kill (2017)} \textcolor{red}{Rogue (2007)}. Excelent movies for you?
\\\hline
\textbf{User:}  & Maybe if it comes on TV  \\\hline
\textbf{Human:} & You should try them. you’re welcome \\
% \textbf{KECRS:} & You too ! \\
\textbf{KGSF:}  & I hope you enjoy them. Have a great day! \\ 
\textbf{Revcore:} & I hope you enjoy them.  \\
\textbf{KLEVER:}  & You're welcome . Bye \\\hline
\end{tabularx}}

\label{tab:case_study}
\vspace{-0.5cm}
\end{table}

\section{Conclusion}
\label{sec:conclusion}

In this paper, we introduce \ourmodel{}, a novel CRS framework that directly handle the misalignment between words and entities by modeling them in the same latent space. We leverage rich textual features (such as item categories, user reviews and entity descriptions) to construct an item descriptive graph which connects two pieces of data signal and devise a knowledge-enhanced fusion module to jointly learn both word and entity representations. Based on the constructed graph, we also introduce a bag-of-words loss to promote the model to generate more informative responses. Extensive experiments demonstrate the superiority of \ourmodel{} compared with state-of-the-art CRS models, especially in the cases that lack information from the user input.

\bibliography{anthology,custom}

\begin{thebibliography}{50}
\expandafter\ifx\csname natexlab\endcsname\relax\def\natexlab#1{#1}\fi

\bibitem[{Arabshahi et~al.(2021)Arabshahi, Lee, Bosselut, Choi, and
  Mitchell}]{arabshahi2021}
Forough Arabshahi, Jennifer Lee, Antoine Bosselut, Yejin Choi, and Tom
  Mitchell. 2021.
\newblock \href {https://doi.org/10.18653/v1/2021.emnlp-main.588}
  {Conversational multi-hop reasoning with neural commonsense knowledge and
  symbolic logic rules}.
\newblock In \emph{Proceedings of the 2021 Conference on Empirical Methods in
  Natural Language Processing}, pages 7404--7418, Online and Punta Cana,
  Dominican Republic. Association for Computational Linguistics.

\bibitem[{Berg et~al.(2017)Berg, Kipf, and Welling}]{gcmc}
Rianne van~den Berg, Thomas~N. Kipf, and Max Welling. 2017.
\newblock \href {https://doi.org/10.48550/ARXIV.1706.02263} {Graph
  convolutional matrix completion}.

\bibitem[{Bizer et~al.(2009)Bizer, Lehmann, Kobilarov, Auer, Becker, Cyganiak,
  and Hellmann}]{dbpedia}
Christian Bizer, Jens Lehmann, Georgi Kobilarov, Sören Auer, Christian Becker,
  Richard Cyganiak, and Sebastian Hellmann. 2009.
\newblock \href {https://doi.org/https://doi.org/10.1016/j.websem.2009.07.002}
  {Dbpedia - a crystallization point for the web of data}.
\newblock \emph{Journal of Web Semantics}, 7(3):154--165.
\newblock The Web of Data.

\bibitem[{Chen et~al.(2019)Chen, Lin, Zhang, Ding, Cen, Yang, and Tang}]{kbrd}
Qibin Chen, Junyang Lin, Yichang Zhang, Ming Ding, Yukuo Cen, Hongxia Yang, and
  Jie Tang. 2019.
\newblock \href {https://doi.org/10.18653/v1/D19-1189} {Towards knowledge-based
  recommender dialog system}.
\newblock In \emph{Proceedings of the 2019 Conference on Empirical Methods in
  Natural Language Processing and the 9th International Joint Conference on
  Natural Language Processing (EMNLP-IJCNLP)}, pages 1803--1813, Hong Kong,
  China. Association for Computational Linguistics.

\bibitem[{Chen et~al.(2020)Chen, Wang, Xie, Parsana, Soni, Ao, and
  Chen}]{explainable}
Zhongxia Chen, Xiting Wang, Xing Xie, Mehul Parsana, Akshay Soni, Xiang Ao, and
  Enhong Chen. 2020.
\newblock \href {https://doi.org/10.24963/ijcai.2020/414} {Towards explainable
  conversational recommendation}.
\newblock In \emph{Proceedings of the Twenty-Ninth International Joint
  Conference on Artificial Intelligence, {IJCAI-20}}, pages 2994--3000.
  International Joint Conferences on Artificial Intelligence Organization.
\newblock Main track.

\bibitem[{Christakopoulou et~al.(2016)Christakopoulou, Radlinski, and
  Hofmann}]{tcrm}
Konstantina Christakopoulou, Filip Radlinski, and Katja Hofmann. 2016.
\newblock \href {https://doi.org/10.1145/2939672.2939746} {Towards
  conversational recommender systems}.
\newblock In \emph{Proceedings of the 22nd ACM SIGKDD International Conference
  on Knowledge Discovery and Data Mining}, KDD '16, page 815–824, New York,
  NY, USA. Association for Computing Machinery.

\bibitem[{Deng et~al.(2021)Deng, Li, Sun, Ding, and Lam}]{unicorn}
Yang Deng, Yaliang Li, Fei Sun, Bolin Ding, and Wai Lam. 2021.
\newblock \href {https://doi.org/10.1145/3404835.3462913} {\emph{Unified
  Conversational Recommendation Policy Learning via Graph-Based Reinforcement
  Learning}}, page 1431–1441. Association for Computing Machinery, New York,
  NY, USA.

\bibitem[{Fleiss and Cohen(1973)}]{kappa}
Joseph~L. Fleiss and Jacob Cohen. 1973.
\newblock \href {https://doi.org/10.1177/001316447303300309} {The equivalence
  of weighted kappa and the intraclass correlation coefficient as measures of
  reliability}.
\newblock \emph{Educational and Psychological Measurement}, 33(3):613--619.

\bibitem[{Gao et~al.(2021)Gao, Lei, He, de~Rijke, and Chua}]{survey}
Chongming Gao, Wenqiang Lei, Xiangnan He, Maarten de~Rijke, and Tat-Seng Chua.
  2021.
\newblock \href {https://doi.org/10.1016/j.aiopen.2021.06.002} {Advances and
  challenges in conversational recommender systems: A survey}.
\newblock \emph{AI Open}, 2:100–126.

\bibitem[{Gu et~al.(2016)Gu, Lu, Li, and Li}]{copy}
Jiatao Gu, Zhengdong Lu, Hang Li, and Victor~O.K. Li. 2016.
\newblock \href {https://doi.org/10.18653/v1/P16-1154} {Incorporating copying
  mechanism in sequence-to-sequence learning}.
\newblock In \emph{Proceedings of the 54th Annual Meeting of the Association
  for Computational Linguistics (Volume 1: Long Papers)}, pages 1631--1640,
  Berlin, Germany. Association for Computational Linguistics.

\bibitem[{Hayati et~al.(2020)Hayati, Kang, Zhu, Shi, and Yu}]{inspired}
Shirley~Anugrah Hayati, Dongyeop Kang, Qingxiaoyang Zhu, Weiyan Shi, and Zhou
  Yu. 2020.
\newblock \href {https://doi.org/10.18653/v1/2020.emnlp-main.654} {{INSPIRED}:
  Toward sociable recommendation dialog systems}.
\newblock In \emph{Proceedings of the 2020 Conference on Empirical Methods in
  Natural Language Processing (EMNLP)}, pages 8142--8152, Online. Association
  for Computational Linguistics.

\bibitem[{He et~al.(2020)He, Deng, Wang, Li, Zhang, and Wang}]{light_gcn}
Xiangnan He, Kuan Deng, Xiang Wang, Yan Li, YongDong Zhang, and Meng Wang.
  2020.
\newblock \href {https://doi.org/10.1145/3397271.3401063} {\emph{LightGCN:
  Simplifying and Powering Graph Convolution Network for Recommendation}}, page
  639–648. Association for Computing Machinery, New York, NY, USA.

\bibitem[{He et~al.(2017)He, Liao, Zhang, Nie, Hu, and Chua}]{ncf}
Xiangnan He, Lizi Liao, Hanwang Zhang, Liqiang Nie, Xia Hu, and Tat-Seng Chua.
  2017.
\newblock \href {https://doi.org/10.1145/3038912.3052569} {Neural collaborative
  filtering}.
\newblock In \emph{Proceedings of the 26th International Conference on World
  Wide Web}, WWW '17, page 173–182, Republic and Canton of Geneva, CHE.
  International World Wide Web Conferences Steering Committee.

\bibitem[{Kang et~al.(2019)Kang, Balakrishnan, Shah, Crook, Boureau, and
  Weston}]{botplay}
Dongyeop Kang, Anusha Balakrishnan, Pararth Shah, Paul Crook, Y-Lan Boureau,
  and Jason Weston. 2019.
\newblock \href {https://doi.org/10.18653/v1/D19-1203} {Recommendation as a
  communication game: Self-supervised bot-play for goal-oriented dialogue}.
\newblock In \emph{Proceedings of the 2019 Conference on Empirical Methods in
  Natural Language Processing and the 9th International Joint Conference on
  Natural Language Processing (EMNLP-IJCNLP)}, pages 1951--1961, Hong Kong,
  China. Association for Computational Linguistics.

\bibitem[{Kipf and Welling(2017)}]{gcn}
Thomas~N. Kipf and Max Welling. 2017.
\newblock Semi-supervised classification with graph convolutional networks.
\newblock In \emph{International Conference on Learning Representations
  (ICLR)}.

\bibitem[{Kong et~al.(2019)Kong, de~Masson~d'Autume, Ling, Yu, Dai, and
  Yogatama}]{mutual}
Lingpeng Kong, Cyprien de~Masson~d'Autume, Wang Ling, Lei Yu, Zihang Dai, and
  Dani Yogatama. 2019.
\newblock \href {http://arxiv.org/abs/1910.08350} {A mutual information
  maximization perspective of language representation learning}.

\bibitem[{Le and Lauw(2021)}]{le2021efficient}
Dung~D Le and Hady Lauw. 2021.
\newblock Efficient retrieval of matrix factorization-based top-k
  recommendations: A survey of recent approaches.
\newblock \emph{Journal of Artificial Intelligence Research}, 70:1441--1479.

\bibitem[{Le and Lauw(2017)}]{le2017indexable}
Dung~D Le and Hady~W Lauw. 2017.
\newblock Indexable bayesian personalized ranking for efficient top-k
  recommendation.
\newblock In \emph{Proceedings of the 2017 ACM on Conference on Information and
  Knowledge Management}, pages 1389--1398.

\bibitem[{Lei et~al.(2020{\natexlab{a}})Lei, He, Miao, Wu, Hong, Kan, and
  Chua}]{ear}
Wenqiang Lei, Xiangnan He, Yisong Miao, Qingyun Wu, Richang Hong, Min-Yen Kan,
  and Tat-Seng Chua. 2020{\natexlab{a}}.
\newblock \href {https://doi.org/10.1145/3336191.3371769}
  {Estimation-action-reflection: Towards deep interaction between
  conversational and recommender systems}.
\newblock \emph{Proceedings of the 13th International Conference on Web Search
  and Data Mining}.

\bibitem[{Lei et~al.(2020{\natexlab{b}})Lei, Zhang, He, Miao, Wang, Chen, and
  Chua}]{scpr}
Wenqiang Lei, Gangyi Zhang, Xiangnan He, Yisong Miao, Xiang Wang, Liang Chen,
  and Tat-Seng Chua. 2020{\natexlab{b}}.
\newblock \href {https://doi.org/10.1145/3394486.3403258} {Interactive path
  reasoning on graph for conversational recommendation}.
\newblock \emph{Proceedings of the 26th ACM SIGKDD International Conference on
  Knowledge Discovery and Data Mining}.

\bibitem[{Li et~al.(2016)Li, Galley, Brockett, Gao, and Dolan}]{diverse}
Jiwei Li, Michel Galley, Chris Brockett, Jianfeng Gao, and Bill Dolan. 2016.
\newblock \href {https://doi.org/10.18653/v1/N16-1014} {A diversity-promoting
  objective function for neural conversation models}.
\newblock In \emph{Proceedings of the 2016 Conference of the North {A}merican
  Chapter of the Association for Computational Linguistics: Human Language
  Technologies}, pages 110--119, San Diego, California. Association for
  Computational Linguistics.

\bibitem[{Li et~al.(2018{\natexlab{a}})Li, Han, and Wu}]{smoothing}
Qimai Li, Zhichao Han, and Xiao-Ming Wu. 2018{\natexlab{a}}.
\newblock \href {https://doi.org/10.48550/ARXIV.1801.07606} {Deeper insights
  into graph convolutional networks for semi-supervised learning}.

\bibitem[{Li et~al.(2018{\natexlab{b}})Li, Ebrahimi~Kahou, Schulz, Michalski,
  Charlin, and Pal}]{redial}
Raymond Li, Samira Ebrahimi~Kahou, Hannes Schulz, Vincent Michalski, Laurent
  Charlin, and Chris Pal. 2018{\natexlab{b}}.
\newblock \href
  {https://proceedings.neurips.cc/paper/2018/file/800de15c79c8d840f4e78d3af937d4d4-Paper.pdf}
  {Towards deep conversational recommendations}.
\newblock In \emph{Advances in Neural Information Processing Systems},
  volume~31. Curran Associates, Inc.

\bibitem[{Li et~al.(2021)Li, Lei, Wu, He, Jiang, and Chua}]{conts}
Shijun Li, Wenqiang Lei, Qingyun Wu, Xiangnan He, Peng Jiang, and Tat-Seng
  Chua. 2021.
\newblock \href {https://doi.org/10.1145/3446427} {Seamlessly unifying
  attributes and items: Conversational recommendation for cold-start users}.
\newblock \emph{ACM Trans. Inf. Syst.}, 39(4).

\bibitem[{Liang et~al.(2021)Liang, Hu, Xu, Miao, He, Chen, Geng, Liang, and
  Jiang}]{ntrd}
Zujie Liang, Huang Hu, Can Xu, Jian Miao, Yingying He, Yining Chen, Xiubo Geng,
  Fan Liang, and Daxin Jiang. 2021.
\newblock \href {https://doi.org/10.18653/v1/2021.emnlp-main.617} {Learning
  neural templates for recommender dialogue system}.
\newblock In \emph{Proceedings of the 2021 Conference on Empirical Methods in
  Natural Language Processing}, pages 7821--7833, Online and Punta Cana,
  Dominican Republic. Association for Computational Linguistics.

\bibitem[{Liao et~al.(2019)Liao, Takanobu, Ma, Yang, Huang, and Chua}]{deepcrs}
Lizi Liao, Ryuichi Takanobu, Yunshan Ma, Xun Yang, Minlie Huang, and Tat-Seng
  Chua. 2019.
\newblock \href {http://arxiv.org/abs/1907.00710} {Deep conversational
  recommender in travel}.

\bibitem[{Liu et~al.(2020)Liu, Wang, Niu, Wu, Che, and Liu}]{multi_type}
Zeming Liu, Haifeng Wang, Zheng-Yu Niu, Hua Wu, Wanxiang Che, and Ting Liu.
  2020.
\newblock \href {https://doi.org/10.18653/v1/2020.acl-main.98} {Towards
  conversational recommendation over multi-type dialogs}.
\newblock In \emph{Proceedings of the 58th Annual Meeting of the Association
  for Computational Linguistics}, pages 1036--1049, Online. Association for
  Computational Linguistics.

\bibitem[{Lu et~al.(2021)Lu, Bao, Song, Ma, Cui, Wu, and He}]{revcore}
Yu~Lu, Junwei Bao, Yan Song, Zichen Ma, Shuguang Cui, Youzheng Wu, and Xiaodong
  He. 2021.
\newblock \href {https://doi.org/10.18653/v1/2021.findings-acl.99}
  {{R}ev{C}ore: Review-augmented conversational recommendation}.
\newblock In \emph{Findings of the Association for Computational Linguistics:
  ACL-IJCNLP 2021}, pages 1161--1173, Online. Association for Computational
  Linguistics.

\bibitem[{Ma et~al.(2021)Ma, Takanobu, and Huang}]{crwalker}
Wenchang Ma, Ryuichi Takanobu, and Minlie Huang. 2021.
\newblock Cr-walker: Tree-structured graph reasoning and dialog acts for
  conversational recommendation.
\newblock In \emph{Proceedings of the 2021 Conference on Empirical Methods in
  Natural Language Processing}, pages 1839--1851. ACL.

\bibitem[{Mikolov et~al.(2013)Mikolov, Sutskever, Chen, Corrado, and
  Dean}]{word2vec}
Tomas Mikolov, Ilya Sutskever, Kai Chen, Greg~S Corrado, and Jeff Dean. 2013.
\newblock \href
  {https://proceedings.neurips.cc/paper/2013/file/9aa42b31882ec039965f3c4923ce901b-Paper.pdf}
  {Distributed representations of words and phrases and their
  compositionality}.
\newblock In \emph{Advances in Neural Information Processing Systems},
  volume~26. Curran Associates, Inc.

\bibitem[{Ren et~al.(2021)Ren, Yin, Chen, Wang, Huang, and Zheng}]{kbqg}
Xuhui Ren, Hongzhi Yin, Tong Chen, Hao Wang, Zi~Huang, and Kai Zheng. 2021.
\newblock \href {https://doi.org/10.1145/3404835.3462839} {Learning to ask
  appropriate questions in conversational recommendation}.
\newblock \emph{Proceedings of the 44th International ACM SIGIR Conference on
  Research and Development in Information Retrieval}.

\bibitem[{Ren et~al.(2020)Ren, Yin, Chen, Wang, Hung, Huang, and Zhang}]{crsal}
Xuhui Ren, Hongzhi Yin, Tong Chen, Hao Wang, Nguyen Quoc~Viet Hung, Zi~Huang,
  and Xiangliang Zhang. 2020.
\newblock \href {https://doi.org/10.1145/3394592} {Crsal: Conversational
  recommender systems with adversarial learning}.
\newblock \emph{ACM Trans. Inf. Syst.}, 38(4).

\bibitem[{Schlichtkrull et~al.(2017)Schlichtkrull, Kipf, Bloem, van~den Berg,
  Titov, and Welling}]{rgcn}
Michael Schlichtkrull, Thomas~N. Kipf, Peter Bloem, Rianne van~den Berg, Ivan
  Titov, and Max Welling. 2017.
\newblock \href {http://arxiv.org/abs/1703.06103} {Modeling relational data
  with graph convolutional networks}.

\bibitem[{Speer et~al.(2017)Speer, Chin, and Havasi}]{conceptnet}
Robyn Speer, Joshua Chin, and Catherine Havasi. 2017.
\newblock Conceptnet 5.5: An open multilingual graph of general knowledge.
\newblock In \emph{Proceedings of the Thirty-First AAAI Conference on
  Artificial Intelligence}, AAAI'17, page 4444–4451. AAAI Press.

\bibitem[{Sun and Zhang(2018)}]{crm}
Yueming Sun and Yi~Zhang. 2018.
\newblock \href {http://arxiv.org/abs/1806.03277} {Conversational recommender
  system}.

\bibitem[{van~der Maaten and Hinton(2008)}]{tsne}
Laurens van~der Maaten and Geoffrey Hinton. 2008.
\newblock \href {http://jmlr.org/papers/v9/vandermaaten08a.html} {Visualizing
  data using t-sne}.
\newblock \emph{Journal of Machine Learning Research}, 9(86):2579--2605.

\bibitem[{Vaswani et~al.(2017)Vaswani, Shazeer, Parmar, Uszkoreit, Jones,
  Gomez, Kaiser, and Polosukhin}]{transformer}
Ashish Vaswani, Noam Shazeer, Niki Parmar, Jakob Uszkoreit, Llion Jones,
  Aidan~N Gomez, \L~ukasz Kaiser, and Illia Polosukhin. 2017.
\newblock \href
  {https://proceedings.neurips.cc/paper/2017/file/3f5ee243547dee91fbd053c1c4a845aa-Paper.pdf}
  {Attention is all you need}.
\newblock In \emph{Advances in Neural Information Processing Systems},
  volume~30. Curran Associates, Inc.

\bibitem[{Veli{\v{c}}kovi{\'{c}} et~al.(2019)Veli{\v{c}}kovi{\'{c}}, Fedus,
  Hamilton, Li{\`{o}}, Bengio, and Hjelm}]{dgi}
Petar Veli{\v{c}}kovi{\'{c}}, William Fedus, William~L. Hamilton, Pietro
  Li{\`{o}}, Yoshua Bengio, and R~Devon Hjelm. 2019.
\newblock \href {https://openreview.net/forum?id=rklz9iAcKQ} {{Deep Graph
  Infomax}}.
\newblock In \emph{International Conference on Learning Representations}.

\bibitem[{Vincent et~al.(2008)Vincent, Larochelle, Bengio, and Manzagol}]{auto}
Pascal Vincent, Hugo Larochelle, Yoshua Bengio, and Pierre-Antoine Manzagol.
  2008.
\newblock \href {https://doi.org/10.1145/1390156.1390294} {Extracting and
  composing robust features with denoising autoencoders}.
\newblock In \emph{Proceedings of the 25th International Conference on Machine
  Learning}, ICML '08, page 1096–1103, New York, NY, USA. Association for
  Computing Machinery.

\bibitem[{Wang et~al.(2019)Wang, He, Wang, Feng, and Chua}]{ngcf}
Xiang Wang, Xiangnan He, Meng Wang, Fuli Feng, and Tat-Seng Chua. 2019.
\newblock \href {https://doi.org/10.1145/3331184.3331267} {Neural graph
  collaborative filtering}.
\newblock In \emph{Proceedings of the 42nd International ACM SIGIR Conference
  on Research and Development in Information Retrieval}, SIGIR'19, page
  165–174, New York, NY, USA. Association for Computing Machinery.

\bibitem[{Xu et~al.(2015)Xu, Wang, Chen, and Li}]{leaky_relu}
Bing Xu, Naiyan Wang, Tianqi Chen, and Mu~Li. 2015.
\newblock \href {http://arxiv.org/abs/1505.00853} {Empirical evaluation of
  rectified activations in convolutional network}.

\bibitem[{Xu et~al.(2020)Xu, Moon, Liu, Liu, Shah, Liu, and Yu}]{memory_graph}
Hu~Xu, Seungwhan Moon, Honglei Liu, Bing Liu, Pararth Shah, Bing Liu, and
  Philip Yu. 2020.
\newblock \href {https://doi.org/10.18653/v1/2020.coling-main.463} {User memory
  reasoning for conversational recommendation}.
\newblock In \emph{Proceedings of the 28th International Conference on
  Computational Linguistics}, pages 5288--5308, Barcelona, Spain (Online).
  International Committee on Computational Linguistics.

\bibitem[{Xu et~al.(2021)Xu, Yang, Xu, Gao, Guo, and Wen}]{adapt}
Kerui Xu, Jingxuan Yang, Jun Xu, Sheng Gao, Jun Guo, and Ji-Rong Wen. 2021.
\newblock \href {https://doi.org/10.1145/3437963.3441791} {Adapting user
  preference to online feedback in multi-round conversational recommendation}.
\newblock In \emph{Proceedings of the 14th ACM International Conference on Web
  Search and Data Mining}, WSDM '21, page 364–372, New York, NY, USA.
  Association for Computing Machinery.

\bibitem[{Zhang et~al.(2019{\natexlab{a}})Zhang, Shi, Zhao, and
  King}]{star_gcn}
Jiani Zhang, Xingjian Shi, Shenglin Zhao, and Irwin King. 2019{\natexlab{a}}.
\newblock \href {https://doi.org/10.48550/ARXIV.1905.13129} {Star-gcn: Stacked
  and reconstructed graph convolutional networks for recommender systems}.

\bibitem[{Zhang et~al.(2019{\natexlab{b}})Zhang, Yao, Sun, and Tay}]{rs_survey}
Shuai Zhang, Lina Yao, Aixin Sun, and Yi~Tay. 2019{\natexlab{b}}.
\newblock \href {https://doi.org/10.1145/3285029} {Deep learning based
  recommender system: A survey and new perspectives}.
\newblock \emph{ACM Comput. Surv.}, 52(1).

\bibitem[{Zhang et~al.(2021)Zhang, Liu, Zhong, Zhang, Wang, and Miao}]{kecrs}
Tong Zhang, Yong Liu, Peixiang Zhong, Chen Zhang, Hao Wang, and Chunyan Miao.
  2021.
\newblock \href {http://arxiv.org/abs/2105.08261} {Kecrs: Towards
  knowledge-enriched conversational recommendation system}.

\bibitem[{Zhang et~al.(2018)Zhang, Chen, Ai, Yang, and Croft}]{saur}
Yongfeng Zhang, Xu~Chen, Qingyao Ai, Liu Yang, and W.~Bruce Croft. 2018.
\newblock \href {https://doi.org/10.1145/3269206.3271776} {Towards
  conversational search and recommendation: System ask, user respond}.
\newblock In \emph{Proceedings of the 27th ACM International Conference on
  Information and Knowledge Management}, CIKM '18, page 177–186, New York,
  NY, USA. Association for Computing Machinery.

\bibitem[{Zhou et~al.(2020{\natexlab{a}})Zhou, Zhao, Bian, Zhou, Wen, and
  Yu}]{kgsf}
Kun Zhou, Wayne~Xin Zhao, Shuqing Bian, Yuanhang Zhou, Ji{-}Rong Wen, and
  Jingsong Yu. 2020{\natexlab{a}}.
\newblock \href {https://dl.acm.org/doi/10.1145/3394486.3403143} {Improving
  conversational recommender systems via knowledge graph based semantic
  fusion}.
\newblock In \emph{{KDD} '20: The 26th {ACM} {SIGKDD} Conference on Knowledge
  Discovery and Data Mining, Virtual Event, CA, USA, August 23-27, 2020}, pages
  1006--1014.

\bibitem[{Zhou et~al.(2020{\natexlab{b}})Zhou, Zhou, Zhao, Wang, and
  Wen}]{tg_redial}
Kun Zhou, Yuanhang Zhou, Wayne~Xin Zhao, Xiaoke Wang, and Ji-Rong Wen.
  2020{\natexlab{b}}.
\newblock \href {https://doi.org/10.18653/v1/2020.coling-main.365} {Towards
  topic-guided conversational recommender system}.
\newblock In \emph{Proceedings of the 28th International Conference on
  Computational Linguistics}, pages 4128--4139, Barcelona, Spain (Online).
  International Committee on Computational Linguistics.

\bibitem[{Zou et~al.(2020)Zou, Chen, and Kanoulas}]{qrec}
Jie Zou, Yifan Chen, and Evangelos Kanoulas. 2020.
\newblock \href {https://doi.org/10.1145/3397271.3401180} {Towards
  question-based recommender systems}.
\newblock \emph{Proceedings of the 43rd International ACM SIGIR Conference on
  Research and Development in Information Retrieval}.

\end{thebibliography}
\bibliographystyle{acl_natbib}

\newpage
\appendix

\section{Appendix}
\label{sec:appendix}

We show some examples of movies and their corresponding top-10 keywords in table \ref{tab:example}.

\begin{table}
\centering
\caption{ Some examples of movies and their corresponding top-10 keywords.}
\begin{tabularx}{\columnwidth}{lX}
\hline
\textbf{Movies} & \textbf{Top-10 Descriptive Words} \\
\hline
\textbf{The Conjuring:}  & conjuring, horror, mystery, thriller, cinema, ghost, exorcism, possession, haunted, supernatural \\

\textbf{It (2017):}  & horror, cinema, entertainment, vertigo, evil, clown, killer, balloon, based, supernatural \\

\textbf{Rings (2017):}  & rings, drama, horror, mystery, romance, image, nation, ghost, film, television \\

\textbf{Interstellar (2014):}  & interstellar, adventure, drama, sci-fi, legendary, entertainment, astronaut, family, gravity, relativity  \\

\textbf{Rogue (2007):}  & rogue, action, drama, horror, thriller, dimension, films, creek, pictures, entertainment, wildlife, travel \\ \hline

\end{tabularx}
\label{tab:example}
\end{table}

\subsection{Statistics of Dataset and Implementation Details}
\label{sec:stats-dataset}

We use Pytorch framework \footnote{\url{https://pytorch.org/}} to implement our CRS framework and train the model on 1GPU NVIDIA A100 40G card. The total training time takes approximately 15 hours. 
The final results are averaged on 5 times with different random seeds. We set the dimensionality of hidden vectors for recommendation engine (including graph convolutional layers) and dialog module to 128 and 300 respectively. The minimal frequency of a words $m$ is 10 and number of keywords $k$ is 30. Besides, the layer number number is 1 for both GNN networks and the Adam optimizer has the learning rate of 1e-3. Total number of training parameters is approximately 150.47 M. \ourmodel{} does not rely on the MIM objective; therefore, it does not require an additional pretraining phase for the MIM loss function to achieve better results. Hence, we first train the recommendation engine and the link prediction loss jointly; beside, we set the weight $\lambda$ of the link prediction loss to 0.25. After training the recommendation engine and the link prediction task, we train the response generation module and set $\lambda_{2}$ to 0.5 for the bag-of-words loss. Code and data are made available as supplementary to foster further research. 

For word and sentence tokenization, we utilize NLTK with default configurations provided by the software. 

\subsection{Instructions for Human Evaluation}
\label{sec:instruct-manual-eval}
Given a conversation context, we ask the annotators to score each response according to some facets ( here we take into account two aspects i.e Fluency and Informativeness). The rating scores must be in range of [1,3] and a higher score is corresponding to a better response. For Fluency, we ask the annotators to check whether the generated responses are grammar-correct and understandable.The defined instructions for Fluency is as follows:
\begin{itemize}
    \item (1): If the sentence is strictly grammar-incorrect and the annotator can not understand the meaning or the intention of the sentence. (e.g: I 'm not sure if it is it , but I have n't seen it 's it 's a big)
    \item (2):  The sentence may have some minor mistakes but it is still understandable. 
    \item (3): The generated sentence is grammar-correct as well as the annotator can comprehend its meaning or intention. (e.g: How about Air Force One  (1997) ?)
\end{itemize}
For Informativeness, the annotator need to check whether the response contains recommended items or relevant information. The defined instruction of Informativeness is as follows.
\begin{itemize}
    \item (1): The generated response does not contain any useful information such as items and entities.
    
    \item (2): The generated response contains either recommended items or related information. 
    
    \item (3): The generated response not only contains relevant recommendations but it also provides some additional related knowledge about the recommended items. 
    
\end{itemize}

\begin{table}
\centering
\caption{ The statistics of Redial dataset and our word-entity graph. “\#" means number and “avg" refers to average. "IDG" denotes for the item descriptive graph.}
\begin{tabularx}{\columnwidth}{ll|ll}
\hline
\textbf{Redial} &  & \textbf{W-E} & \\
\hline
\# of convs & 10006 & \# of movies & 5879  \\
\# of utterances  & 182150 & \# of words & 10575 \\
\# of users & 956 & \# of edges & 313606 \\
\# of movies & 6924 & \ \_ & \_  \\
% avg token length & 6.8 & \ \_ & \_  \\
avg turn \# & 18.2 & \ \_ & \_  \\\hline
\end{tabularx}
\label{tab:dataset}
\end{table}

\subsection{Additional Experimental Results}
\label{sec:exp-num-des-words}

\paragraph{\bf{Number of descriptive words $k$:}} We  conduct an experiment to investigate the research question: \textit{How the number of descriptive words $k$ for each entity in the item descriptive graph effects to our model performance ?}  Table \ref{tab:kg_study} reports the results with different values of $k$. Apparently, it is insufficient for the model to produce meaningful representations when using too few contextual words, i.e. entities are only aligned with a very small number of indicative terms. In contrast, utilizing too many words leads to noisy alignments and may degrade the model performance. This is consistent with our assumption in the case of the KGSF model which aligns all words and entities co-occurring in a conversation. The result shows that \ourmodel{} achieves the highest score when $k = 30$.

\begin{table}
\centering
\caption{ Recommendation performance of \ourmodel{} with different number of keywords $k$.}
\begin{tabular}{lc|c|c|c}
\hline
\textbf{Num of words $k$} & \textbf{R@1} & \textbf{R@10} & \textbf{R@50}\\
\hline
10  & 0.036 & 0.196 & 0.369  \\
20  & 0.039 & 0.196 & 0.373  \\ 
30  & \textbf{0.041} & \textbf{0.203} & \textbf{0.383}  \\
40  & 0.037 & 0.198 & 0.371  \\\hline
\end{tabular}

\label{tab:kg_study}
\end{table}

\subsection{Additional Cold-Start Examples}
\label{sec:add-coldstart-examples}

Table \ref{tab:cold_start} shows some additional cold-start conversations. For convenient, we represent them in form of a historical context and its corresponding responses from different agents including Human, KGSF, Revcore and Our model respectively. We mark all the mentioned items in blue color and the user preferences in red color for easy
reading. 

\begin{table*}
\centering
\caption{ Some cold-start conversation examples on movie item recommendation. The responses of KGSF, Revcore and KLEVER are from the test results for corresponding models. Human responses are ground-truth. <||> denotes to switch conversation roles.}
\begin{tabularx}{\textwidth}{lX}
\hline
& \textbf{Conversation Example} \\
\hline
\textbf{Context 1:}  & Hi How are you ?  <||>  Hello there How are you ? Tell me what kind of movies would you like to watch ?  <||>  \textcolor{blue}{Romance} and \textcolor{blue}{comedy}\\\hline 
\textbf{Human:} & I can recommend \textcolor{red}{Click (2010)} by Adan Sandler. You can smile a lot there !\\ 
\textbf{KGSF:}  & I would recommend some movies for you ? \\ 
\textbf{Revcore:} & Hi , what kind of movies do you like ?   \\
\textbf{KLEVER:}  & I'm doing great . How about \textcolor{red}{How to Lose a Guy in 10 Days (2003)} ? It 's a \textcolor{blue}{comedy} \\\hline
\textbf{Context 2:}  & Hello ! Do you like \textcolor{blue}{classic} movies ?  <||>  Not so much . I like \textcolor{blue}{action} movies from the \textcolor{blue}{90s} . Seen anything good like that ?\\\hline
\textbf{Human:} & Yes great decade for \textcolor{blue}{actions} ! \\
\textbf{KGSF:}  & I would recommend some movies for you ? \\ 
\textbf{Revcore:} & What kind of movies do you like ? \\
\textbf{KLEVER:}  & What about \textcolor{red}{The Terminator (1984)} ?\\\hline
\textbf{Context 3:}  & Hi How are you ?  <||>  Hello I'm awesome and you ?  <||>  I’ m great ! Thanks for asking what kind of movies you like  <||>  I'm open to any option I like \textcolor{blue}{Comedy} and \textcolor{blue}{Action} movies\\\hline
\textbf{Human:} & Have you seen \textcolor{red}{Black Panther (2018)} \textcolor{blue}{Love comedy} ! !\\
\textbf{KGSF:}  & I would recommend some movies for you ? \\ 
\textbf{Revcore:} &  What kind of movies do you like ? \\
\textbf{KLEVER:}  &  I would recommend \textcolor{red}{The Big Sick (2017)}\\\hline
\textbf{Context 4:}  & Hello , how are you ? Hope you 're having a great day .  <||>  I 'm great thank you ! I like \textcolor{blue}{romantic comedies} . Do you have any good ideas for movies ? \\\hline
\textbf{Human:} & One of my favorites is \textcolor{red}{Overboard  (1987)} its a \textcolor{blue}{classic} and the main characters were actually married in \textcolor{blue}{real life} so it was \textcolor{blue}{fun} to see them acting together . \\
\textbf{KGSF:}  & I would recommend some movies to you like to watch ?  \\ 
\textbf{Revcore:} & Hi ! I 'm sure you will like to watch some of my favorites .  \\
\textbf{KLEVER:}  &  I 'm doing great. How about \textcolor{red}{Meet the Fockers (2004)} ?\\\hline
\textbf{Context 5:}  & Hello i need a movie for watch with my \textcolor{blue}{family} can you help me please\\\hline
\textbf{Human:} & \textcolor{red}{The Lion King (1994)} is a good choice \\
\textbf{KGSF:}  & Sure can  \\ 
\textbf{Revcore:} & What type of movies do you like ?  \\
\textbf{KLEVER:}  &  Hello ! What about \textcolor{red}{Toy Story (1995)} ? ?\\\hline
\textbf{Context 6:}  & Hello  <||>  Can you give me any movie sugestions ?  <||>  tell me what kind of movies do you like ?  <||>  I like \textcolor{blue}{comedy} movies\\\hline
\textbf{Human:} & i would recommend you \textcolor{red}{Click (2006)} it is very \textcolor{blue}{funny} i also would recommend you \textcolor{red}{You Don't Mess with the Zohan (2008)}\\
\textbf{KGSF:}  & I would recommend some for you ? \\ 
\textbf{Revcore:} & What kind of movies do you like ? \\
\textbf{KLEVER:}  & Hi ! What about \textcolor{red}{How to Lose a Guy in 10 Days (2003)} ? \\ \hline
\end{tabularx}
\label{tab:cold_start}
\end{table*}

\end{document}